\documentclass[prd,nofootinbib,superscriptaddress]{revtex4}
\usepackage{epsfig}
\usepackage{amsmath}
\usepackage{amsfonts}
\usepackage{amssymb}
\usepackage{float}
\usepackage{color}
\usepackage{graphicx}
\usepackage{graphics}
\usepackage{hyperref}
\hypersetup{}
\usepackage{epstopdf}
\usepackage{multirow}

\definecolor{rosso}{cmyk}{0,1,1,0.4}
\definecolor{rossos}{cmyk}{0,1,1,0.55}
\definecolor{rossoc}{cmyk}{0,1,1,0.2}
\definecolor{blu}{cmyk}{1,1,0,0.3}
\definecolor{blus}{cmyk}{1,1,0,0.6}
\definecolor{bluc}{cmyk}{1,1,0,0.1}
\definecolor{verde}{cmyk}{0.92,0,0.59,0.25}
\definecolor{verdec}{cmyk}{0.92,0,0.59,0.15}
\definecolor{verdes}{cmyk}{0.92,0,0.59,0.4}

\begin{document}

\title{Searching for Heavy Charged Higgs Bosons through Top Quark Polarization}

\author{Abdesslam Arhrib}
\email{aarhrib@gmail.com}
\affiliation{D\'{e}partement de Math\'{e}matiques, 
Facult\'{e} des Sciences et Techniques,
Universit\'{e} Abdelmalek Essaadi, B. 416, Tangier, Morocco.}

\author{Adil Jueid}
\email{adil.jueid@sjtu.edu.cn}

\affiliation{INPAC, Shanghai Key Laboratory for Particle Physics and Cosmology,
Department of Physics and Astronomy, Shanghai Jiao Tong University,
Shanghai 200240, China.}

\author{Stefano Moretti}
\email{s.moretti@soton.ac.uk}
\affiliation{School of Physics and Astronomy, University of Southampton,\\
Southampton, SO17 1BJ, United Kingdom.}

\begin{abstract}
We study the production of a heavy charged Higgs boson at the Large Hadron Collider (LHC) in $gb \to H^- t$ within a 2-Higgs Doublet Model (2HDM). 
The chiral structure of the $H^+ \bar{t} b$ coupling can trigger a particular spin state of the top quark produced
 in the decay of a charged Higgs boson and, therefore, is sensitive to the underlying mechanism of the Electro-Weak Symmetry 
 Breaking (EWSB). Taking two benchmark models (2HDM type-I and 2HDM type-Y) as an example, we show that inclusive 
 rates, differential distributions as well as forward-backward asymmetries of the top quark’s 
 decay products can be used to search for heavy charged Higgs bosons as well as a model discriminators.
\end{abstract}
\maketitle

\section{Introduction}	

The top quark is believed to play an important role in  new phenomena beyond the Standard Model (SM).
It was discovered at the  Tevatron by the D0 \cite{Abachi:1995iq} 
and CDF \cite{Abe:1995hr} collaborations about two decades ago. It is the heaviest
elementary particle we know of with a mass close to the EW scale ($m_t = 172.5$ GeV). The implications of this are manifold. (i) The top quark has a very short lifetime compared to hadronization time scales. (ii) It strongly couples to the SM Higgs boson and therefore can affect its phenomenology in a significant way. (iii) Its mass is an important input for  studies related to the vacuum stability of the universe. However, the large mass of the top quark also implies that it can be produced in reasonable numbers only at high energy accelerators, such as the Large Hadron Collider (LHC). Further, the short lifetime of the top quark ($\tau_t \simeq G_F^{-1} m_t^{-3} \ll m_t/\Lambda_\textrm{QCD}^2$) implies that all its fundamental properties can be pinned down by studying its decay products (for a review, see, e.g., \cite{Bernreuther:2008ju} and references therein). 

The 2-Higgs Doublet Model (2HDM) is one of the  simplest extensions of the SM proposed four decades ago (for a review, see, e.g.,  \cite{Branco:2011iw, Akeroyd:2016ymd}). In this model, two complex doublets 
are introduced to break the EW gauge symmetry and give rise to fermion and gauge boson masses. After Electroweak Symmetry Breaking (EWSB), there are five physical Higgs states (which can be regarded as mediators of mass generation dynamics): two neutral CP-even ones ($h^0$ and $H^0$, with $m_{h^0}<m_{H^0}$), one CP-odd neutral one ($A^0$) and a pair of charged ones ($H^\pm$). The top quark couples to all of these particles. Hence, top quark processes have been studied extensively within the 2HDM  because  the structure of the new Yukawa couplings can  reveal the properties of the underlying 2HDM \cite{Stange:1993td, Zhou:1996dx, Hollik:1997hm, Denner:1992vz, Bernreuther:2008us, Huitu:2010ad,  Arhrib:2016vts, Eilam:1990zc}.

The purpose of this short report is to discuss the polarization effects of the top quark in studies of the charged Higgs bosons at the LHC taking the 2HDM as a benchmark model. We discuss two  Yukawa realizations of the 2HDM:  i.e.,  the 2HDM Type-I (2HDM-I) and  2HDM Type-Y (2HDM-Y). The spin observables (which consist of angular and energy distributions of the decay products) of the top quark can improve the sensitivity of searches for charged Higgs bosons so long that the chiral structure of the ${H^+ \bar{t} b}$ vertex is non-trivial.

\section{Production of charged Higgs bosons in association with tops}

\begin{figure}[!t]
 \centering
 \includegraphics[width=0.48\linewidth]{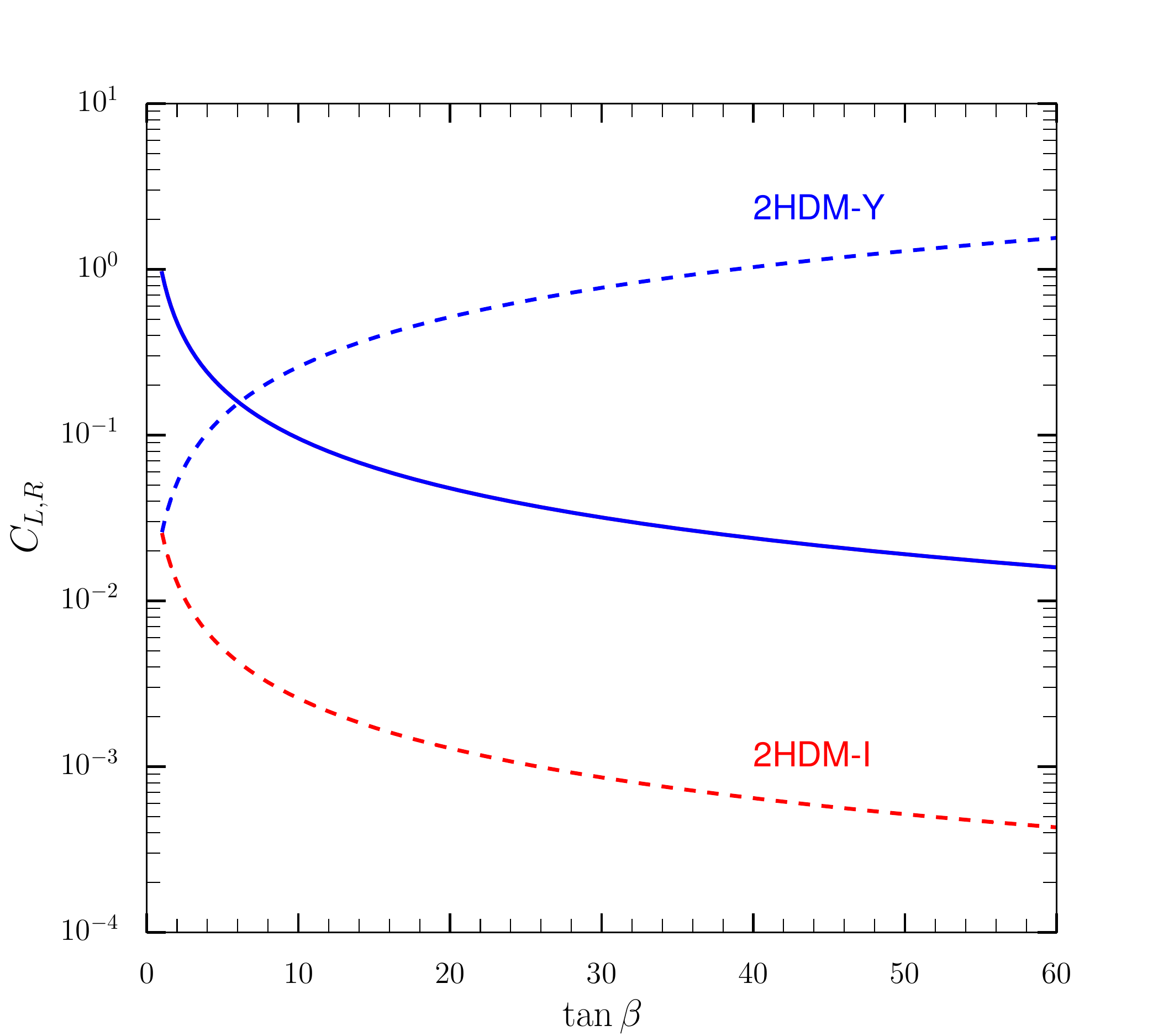}
 \hfill
 \includegraphics[width=0.48\linewidth]{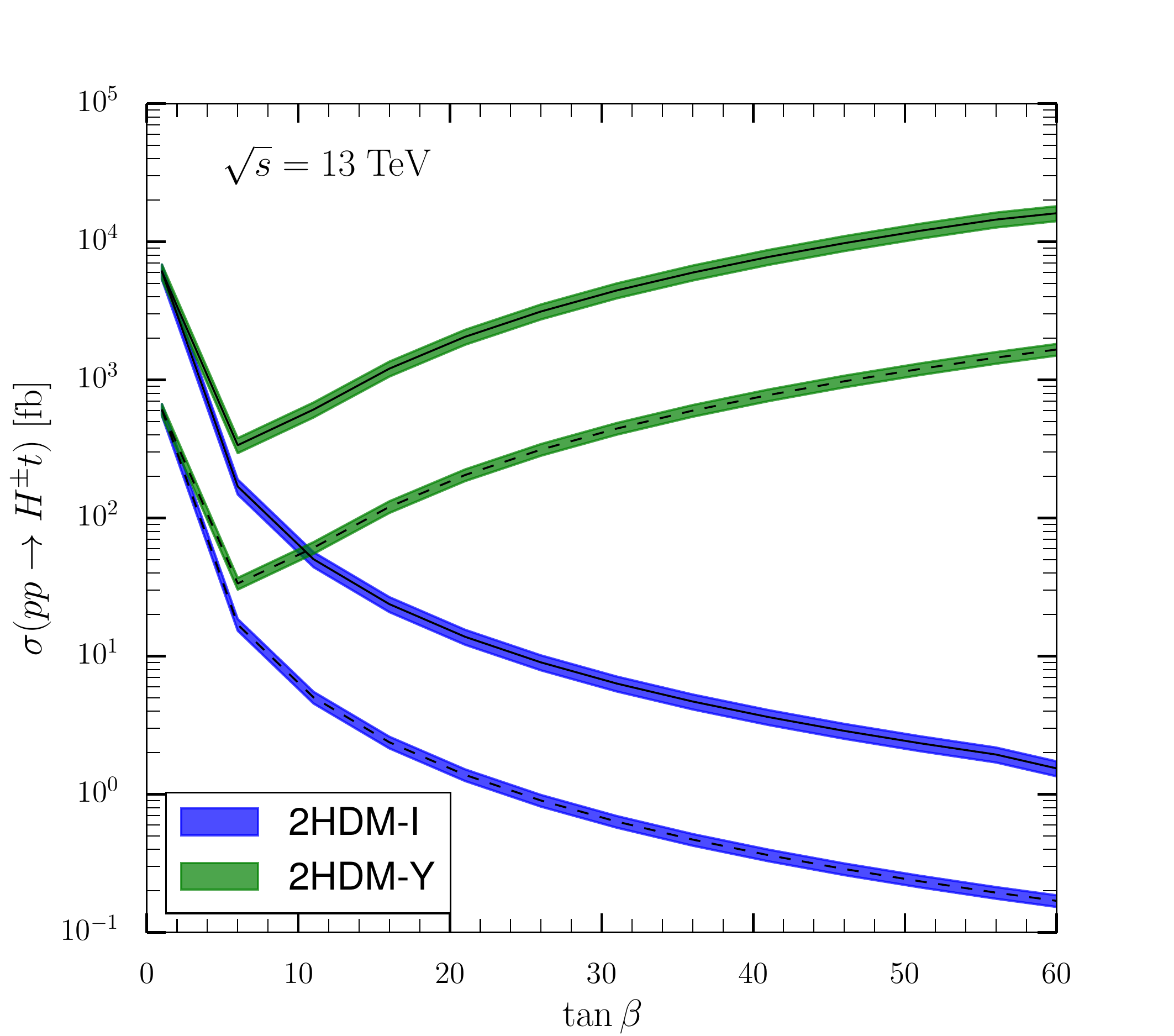}
 \caption{
 \emph{Left}: The chiral form factors of the $H^+\bar tb$ vertex as a function of $\tan\beta$ for the 2HDM-I (red) and 2HDM-Y (blue), where the overlapping solid(separate dashed) lines show the$ L(R)$-handed components. 
 \emph{Right}: The $H^\pm $ production cross section as a function of $\tan\beta$ in the 2HDM-I (blue) and 2HDM-II (green), where the 
 solid(dashed) lines show the cross sections for $m_{H^\pm} = 200~  \textrm{GeV}$($m_{H^\pm} = 500 \textrm{ GeV}$).}
\label{xsec-tHch}
\end{figure}

For details about the model and the constraints used in this study, we refer the interested reader to \cite{Arhrib:2018bxc}. It is customary to study $H^\pm$ production in association with a top quark ($bg\to tH^- $ + c.c.) \cite{Akeroyd:2016ymd} so that the cross section is controlled by the ${H^+\bar{t} b}$ coupling. It is  convenient to write the aforementioned vertex in the form
\begin{eqnarray}
 g_{\bar{t} bH^+} = i \left(C_L P_L + C_R P_R \right), 
 \label{Htb-coupling}
\end{eqnarray}
with $C_L = \frac{1}{\sqrt{2} v} m_t \kappa_u^A$ and 
$C_R = \frac{1}{\sqrt{2} v} m_b \kappa_d^A$, wherein $\kappa_{u,d}^A$ are the Yukawa couplings (please see, e.g., \cite{Aoki:2009ha} from more details about the phenomenology of the Yukawa sector in the 2HDM). The nature of the chiral structure of the $H^+ \bar t b$ coupling depends on $\tan\beta$. First, in the 2HDM-I (and 2HDM-X), both the $R$- and $L$-handed components are proportional to $1/\tan\beta$ and hence, given that $C_L \propto m_t$ while $C_R \propto m_b$, the $H^+\bar t b$ coupling is dominated by the $L$-handed component. In contrast, in the 2HDM-II (and 2HDM-Y), the $L$- and $R$-handed components of  the $H^+\bar t b$ coupling behave differently depending on $\tan\beta$: (i) for $\tan\beta < \sqrt{m_t/m_b}$($\tan\beta>\sqrt{m_t/m_b}$) the coupling is dominated by the $L(R)$-handed component;  (ii) for $\tan\beta\approx \sqrt{m_t/m_b}$, the coupling is purely scalar (with no $\gamma_5$ structure). 
We display in Fig. \ref{xsec-tHch} the dependence of the chiral components $C_{L,R}$ of the $H^+\bar  tb$ vertex and the production cross section upon $\tan\beta$. We can see that the cross section in the 2HDM-Y falls to a dip around  $\tan^2\beta\approx m_t/m_b$  and then increases for large $\tan\beta$. In the 2HDM-I, however, it decreases always for increasing $\tan\beta$. As for the $H^+\bar tb$ vertex,  we notice first that $C_L$ is decreasing as a function of $\tan\beta$  in both types of 2HDM (the corresponding lines in fact overlap). However, in the 2HDM-Y, $C_R$ can be about two orders of magnitude larger than $C_L$ for large $\tan\beta$ values. 

For definiteness, in what follows,  we will choose our benchmark scenarios with $\tan\beta=1$ for the 2HDM-I and $\tan\beta=50$ 
 for the 2HDM-Y which give:
\begin{itemize}
    \item $(C_L,C_R) = (0.94, -0.025)$ for 2HDM-I;
    \item $(C_L,C_R) = (0.019, 1.3)$ for 2HDM-Y.
\end{itemize}
 Before closing this section, we briefly discuss Next-to-Leading Order (NLO) QCD corrections to the production cross section. It was shown that corrections to kinematical quantities (such as the $b$-jet transverse momentum) imply an almost constant $K$-factor \cite{Degrande:2015vpa}. These corrections also improve the agreement between the 4-Flavor scheme (4FS) and 5-Flavor Scheme (5FS)\footnote{We notice that the $gg\to \bar b t H^-$ process is the 4FS equivalent of the $bg\to tH^-$ channel in the 5FS \cite{Guchait:2001pi}.}. The authors of \cite{Godbole:2011vw} have found that spin observables used in this paper are mildly dependent on the perturbative order used in the calculations: specifically, \emph{(i)} no substantial corrections to angular observables and \emph{(ii)} constant corrections to observables based on the $b$-jet energy distribution. 

\section{Phenomenological Setup}

In this section, we discuss the observables that we have used to study the sensitivity of this production process to
top polarization effects, as a function of $\tan\beta$ and $m_{H^\pm}$. We believe that these observables are important to pursue two goals: \emph{(i)} distinguish the signal from the SM backgrounds and \emph{(ii)} disentangle different Yukawa realizations of the 2HDM from each other (or different BSM models from each other).
 
First, one can study the 
 differential distribution 
 in $\cos\theta_\ell^a$ of the emerging lepton which is $100\%$ correlated to the top quark producing it
\begin{equation}
   \frac{1}{\sigma} \frac{\text{d}\sigma}{\text{d}\cos\theta_\ell^a} = 
   \frac{1}{2}\bigg(1 + \alpha_{\ell^\pm} P_{t,\bar{t}} \cos\theta_{\ell}^a\bigg),
 \label{theta}
 \end{equation}
wherein $\alpha_{\ell^\pm}$ is the so-called spin analyzing power of the  charged lepton and $\theta_{\ell^a} = \measuredangle (\hat{\ell}^\pm, \hat{S}_a)$, with $\hat{\ell}^\pm$ being the direction of flight of the charged lepton in the top quark rest frame and $\hat{S}_a$ the spin quantization axis in the basis $a$. In this work, we focus on the helicity basis where the spin quantization axis is defined to be the direction of motion of the top quark in the so-called ($t\bar{t})$ Zero Momentum Frame (ZMF). It was found that energy distributions of the decay products (and their ratios) in the laboratory frame are excellent probes of top quark polarization. These observables were proposed by \cite{Shelton:2008nq} as a probe of new physics and used as a way to probe anomalous $W^+\bar tb$ couplings in \cite{Prasath:2014mfa, Jueid:2018wnj}. They are given by 
\begin{equation}\label{eq:u_and_z_and_xl}
u = \frac{E_\ell}{E_\ell+E_b}, \qquad
z = \frac{E_b}{E_t}, \qquad
 x_\ell = \frac{2 E_\ell}{m_t},
\end{equation}
where $E_\ell$, $E_b$ and $E_t$ are the energies of the charged lepton, 
$b$-jet and  the top quark in the $pp$ center-of-mass frame, respectively. 

We use \textsc{Madgraph5\_aMC@NLO} \cite{Alwall:2011uj}, \textsc{MadSpin} \cite{Artoisenet:2012st} and \textsc{Pythia8} \cite{Sjostrand:2014zea} for the generation of the hard-scattering processes at LO, decay of heavy-resonances and showering/hadronization, respectively. Then, \textsc{Rivet} 2.5.4 \cite{Buckley:2010ar} was used for particle-level analysis where jets are clustered with the help of \textsc{FastJets} \cite{Cacciari:2011ma} using the  anti-$k_\perp$ algorithm with jet radius $D=0.5$. Top quark candidates were reconstructed using the \textsc{PseudoTop} definition \cite{Collaboration:2267573} used widely by the ATLAS and CMS collaborations. We finally use a \textsc{Rivet} analysis for the validation of the CMS measurement of the $t\bar{t}$ differential cross section at $\sqrt{s}=8$ TeV \cite{Khachatryan:2015oqa}. 

\section{Results}

Events are selected if they contain exactly one isolated charged lepton (electron or muon not from tau decays), at least $5$ jets (where at least $3$ of these are $b$-tagged) and missing transverse energy (which corresponds to the SM neutrinos from $W^\pm$ boson decays). We require the presence of one electron(muon) with $p_T > 30$ GeV($p_T > 27$ GeV) and $|\eta|< 2.5$($|\eta| < 2.4$. The missing transverse energy  is required to satisfy $E_T^\textrm{miss} > 20$ GeV. We first require $p_T > 30$ GeV and $|\eta| < 2.4$ for all the jets in an event. We then refine our selection criteria by vetoing events which do not have a leading jet with $p_T > 50$ GeV.  This set of cuts will be denoted by \textsc{Cuts1}. We impose two additional cuts, denoted by \textsc{Cuts2}(\textsc{Cuts3}) on the scalar sum of the jet transverse momenta: i.e. we impose $H_T > 500$~GeV($1000$ GeV). 

For the production of a charged Higgs boson in association with a top quark followed by the $H^\pm \to t b$ decay, where one top decays hadronically and the other one leptonically, there are many background contributions. The most important ones are the exclusive production of a top (anti)quark pair in association with a $b$-quark (i.e., $t\bar t b$ + c.c.),  {top quark pair production in association with a light jet} and $t\bar{t}$ inclusive production. The first one is completely irreducible while the second one {might contribute since the associated light jet can be mis-tagged as a $b$-jet}, then, the third background is partially reducible since the production of additional $b$-quarks is possible from the parton shower, notably in $g\to b\bar{b}$ splitting, but it is not a leading effect. There are further possible background processes, such as single top, di-boson and $W^\pm$ + jet production, but these are generally negligible compared to the previous ones. 

Using  a standard top (anti)quark reconstruction procedure on both heavy flavor states combined with  the requirements on jet activity and the cuts in $H_T$ will reduce substantially the background. To enable the possible observation of a signal, we compute its significance defined as \cite{Cowan:2010js}
\begin{eqnarray}
\mathcal{S} = \sqrt{2 \bigg((N_s+N_b) \log\bigg(1+\frac{N_s}{N_b}\bigg) - N_s\bigg)},
\end{eqnarray}
where $N_s$($N_b$) is the number of signal(background) events 
after a given selection. We finally compute these for $\mathcal{L} = 200$ and $1000$ fb$^{-1}$ of  integrated LHC luminosity. The obtained values are  displayed in Tab. \ref{table:significance}. 

\begin{table}[!t]
\caption{Signal significance of the two types of 2HDM that we study for $m_{H^\pm}=300, 400~ \text{and}~ 
500$ GeV in the 2HDM-I and $m_{H^\pm}=500, 600~ \text{and}~ 700$ GeV in the 2HDM-Y. The numbers outside(inside) the brackets refer to the case of $200(1000)$ fb$^{-1}$ of total luminosity. Normalization is according to the matched flavor scheme discussed. }
{\begin{tabular}{@{}c  ccc  ccc@{} }
\toprule
 & \multicolumn{6}{c}{$\mathcal{L}=200~(1000)$ fb$^{-1}$} \\
 \toprule
& \multicolumn{3}{c}{\textsc{2HDM-I}} &
\multicolumn{3}{c}{\textsc{2HDM-Y}} \\ 
\toprule
$m_{H^\pm}$ [GeV] &   $300$   & $400$ & $500$ &  $500$ & $600$ & $700$ \\ \hline 
\toprule
Initial events &   $12.63~(28.24)$ &   $5.94~(13.29)$ & $3.01~(6.74)$ &     $5.71~(12.77)$  & $3.03~(6.77)$ &  $1.68~(3.75)$      \\ 
\textsc{Cuts1}   & $37.42~(83.32)$ & $20.41~(45.36)$ & $11.30~(25.32)$ &    $19.15~(43.14)$ &  $10.93~(24.68)$ &  $6.43~(14.31)$   \\
\textsc{Cuts2}   & $25.92~(57.67)$  &  $18.61~(41.47)$ &   $12.36~(27.82)$ &    $20.53~(46.21)$ &  $12.80~(28.82)$ &   $7.78~(17.31)$  \\
\textsc{Cuts3}   & $10.53~(22.83)$  &  $8.12~(17.64)$ & $6.16~(13.62)$ &  $10.00~(22.55)$ &  $7.71~(17.66)$ & $5.87~(13.26)$  \\
\botrule
\end{tabular} \label{table:significance}}
\end{table}

\begin{figure}[!h]
    \centering
    \includegraphics[width=0.48\linewidth]{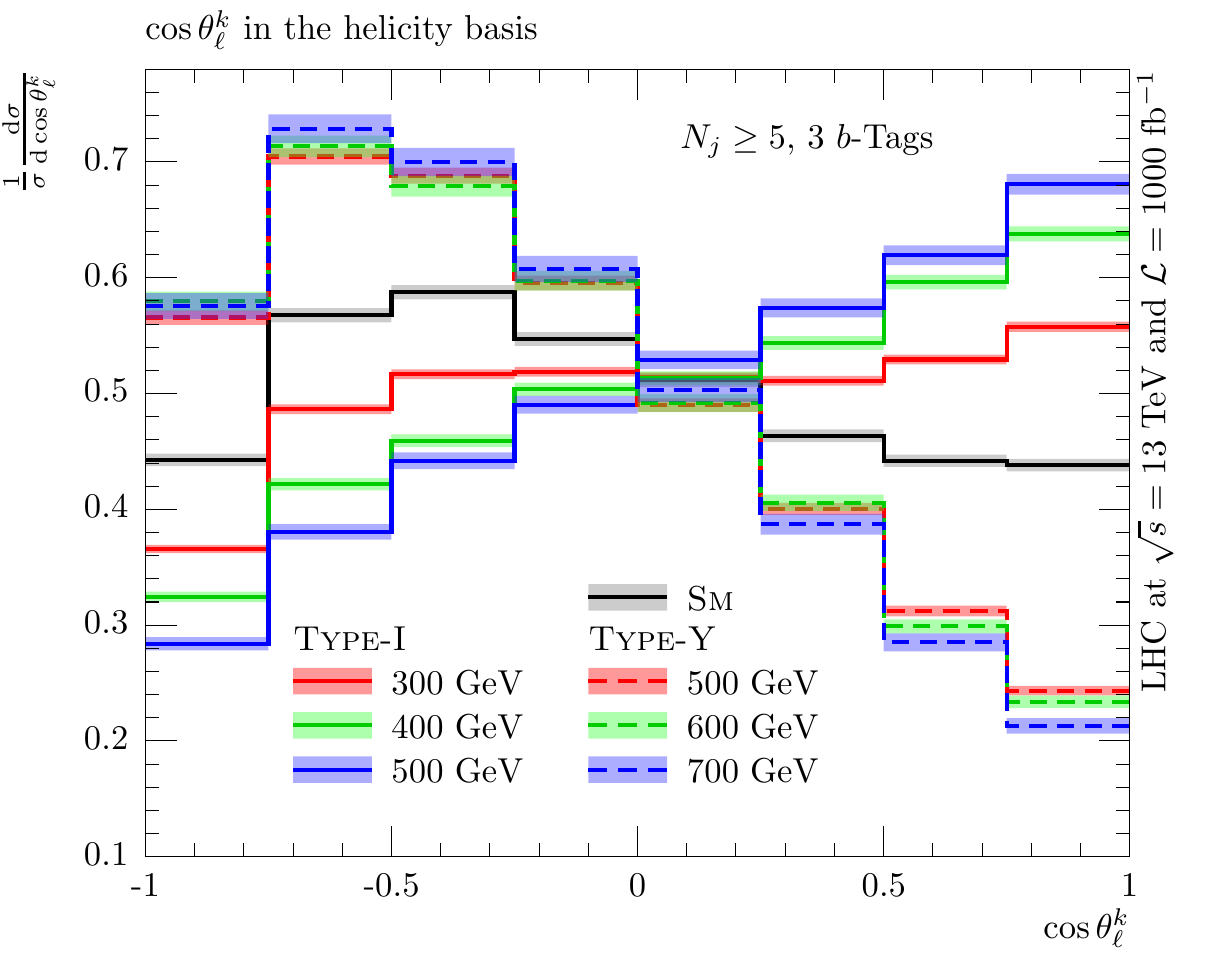}
    \hfill
  \includegraphics[width=0.48\linewidth]{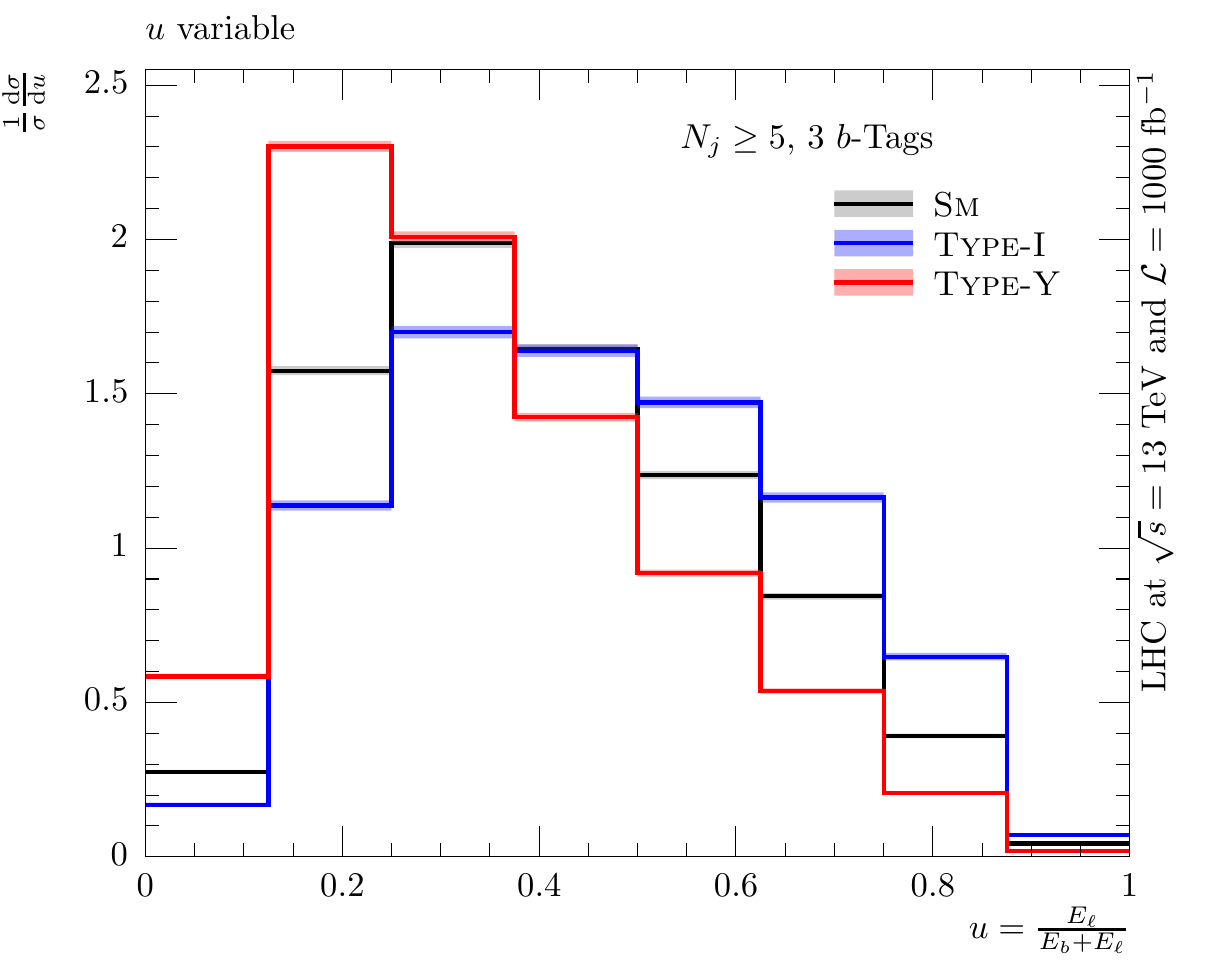}
    \caption{\emph{Left}: The $\cos\theta_\ell^k$ distributions for the SM (solid black), 2HDM-I (solid red, green and blue) and 2HDM-Y (dashed red, green and blue). \emph{Right}: The $u$ distributions for the SM (black), 2HDM-I (blue) and 2HDM-Y (red). Data are for $\sqrt s=13$ TeV and ${\cal L}=1000$ fb$^{-1}$. The shading represents the statistical MC uncertainty on the distributions. In the right panel, the distributions for 2HDM-I and 2HDM-Y correspond to $m_{H^\pm}=500$ GeV. Results are shown after applying the basic selections denoted by \textsc{Cuts1}.}
    \label{fig:fig1}
\end{figure}

\begin{table}[!t]
\caption{The asymmetries $A_{\theta_\ell}, A_{x_\ell} \textrm{ and } A_u$ for the SM background,  2HDM-I
and 2HDM-Y. For each asymmetry, the first row corresponds to the values computed after basic selections while the second row corresponds to the values of $A_X$ after requiring $H_T > 1000$ GeV. The quoted errors for each asymmetry correspond to 
statistical uncertainties.}
{\begin{tabular}{@{}c  c  ccc  ccc@{}}
\toprule
Asymmetry  &  \textsc{Background} &  \multicolumn{3}{c}{\textsc{2HDM-I}} &
\multicolumn{3}{c}{\textsc{2HDM-Y}} \\ 
 &  &  $300$ GeV  & $400$ GeV & $500$ GeV &  $500$ GeV & $600$ GeV & $700$ GeV \\ \hline
$A_{\theta_\ell}$  &  $-0.04\pm0.001$   &     $0.05\pm0.003$    &   $0.14\pm0.004$     &     $0.20\pm0.005$    &    $-0.27\pm0.004$  &   $-0.28\pm0.005$    &  $-0.31\pm0.007$ \\ 
			\cline{2-8}
		       &  $-0.01\pm0.003$ & $0.01\pm0.014$ & $0.08\pm0.012$ & $0.13\pm0.013$ & $-0.28\pm0.009$ & $-0.28\pm0.011$ & $-0.31\pm0.013$ \\ \hline
$A_{x_\ell}$.   &  $0.37\pm0.001$   &     $0.40\pm0.003$    &   $0.52\pm0.003$     &    $0.65\pm0.004$    &    $0.21\pm0.004$  &   $0.27\pm0.005$    &  $0.33\pm0.007$ \\ 
			\cline{2-8}
	               &  $0.54\pm0.003$ & $0.53\pm0.008$ & $0.57\pm0.009$ & $0.65\pm0.010$ & $0.30\pm0.009$ & $0.33\pm0.010$ & $0.38\pm0.012$ \\ \hline
$A_{u}$          &  $-0.35\pm0.001$   &     $-0.30\pm0.003$    &   $-0.22\pm0.004$     &     $-0.16\pm0.005$    &    $-0.58\pm0.003$  &   $-0.58\pm0.004$    &  $-0.58\pm0.006$ \\ 
			\cline{2-8}
		       & $-0.35\pm0.003$ & $-0.27\pm0.009$ & $-0.31\pm0.011$ & $-0.26\pm0.012$ & $-0.63\pm0.008$ & $-0.64\pm0.009$ & $-0.62\pm0.010$  \\ \botrule
\end{tabular} \label{table:asymmetry}}
\end{table}   

In the left panel of Fig. \ref{fig:fig1}, we show the $\cos\theta_\ell^k$ spectrum for the irreducible SM background (denoted by \textsc{Sm}) as well as the 2HDM-I and 2HDM-Y after applying the basic selections. We can see clearly that the \textsc{Sm} curves exhibit almost no dependence on $\cos\theta_\ell^k$ except for some negative values of it. The interesting observation is that the 2HDM-I and 2HDM-Y have opposite slopes and hence different polarization with different sign. We can see that, in 2HDM-Y, the $\cos\theta_\ell^k$ is not able to distinguish between the different masses of the charged Higgs. The $u$-variable is displayed in the right panel of Fig. \ref{fig:fig1} for the \textsc{Sm} plus the 2HDM-I and 2HDM-Y for $m_{H^\pm} = 500~\textrm{GeV}$. We can see that $u$ is more sensitive and can efficiently separate between the three different models\footnote{For more details about the results, please see \cite{Arhrib:2018bxc}.}. 

To quantify the sensitivity of the spin observables on the modelling paradigm, we suggest the use 
of forward-backward asymmetries constructed from the observables defined in Eqs. (\ref{theta})--(\ref{eq:u_and_z_and_xl}). Asymmetries are  resilient to NLO QCD corrections and  the choice of flavor scheme. We define an asymmetry $A_X$ as 
\begin{eqnarray}
A_X = \frac{\sigma(X > X_c) - \sigma(X < X_c)}{\sigma(X > X_c) + \sigma(X < X_c)},
\end{eqnarray}
with $X = \cos\theta_\ell^k, x_\ell, u$ and where $X_c$ is a reference point for the asymmetry $A_X$. In the present study, we choose
the following references points: $\cos\theta_{\ell,c}^k = 0, x_{\ell,c} = 0.6$ and $u_c = 0.5$. In Tab. \ref{table:asymmetry}, we show the values of the three 
asymmetries in the SM and the two usual 2HDM types. It is clear that $A_{\theta_\ell}$ can distinguish between the SM and the two realizations of the 
2HDM considered here. Furthermore, this asymmetry can even distinguish between 
the different masses for the 2HDM-I case. Further, for the 2HDM-Y,  $A_{\theta_\ell}$ is insensitive to the mass of the charged Higgs boson. However, 
$A_{x_\ell}$ is able to remove this degeneracy for the 2HDM-Y.

\section{Conclusions}
We studied the sensitivity to spin observables in charged Higgs boson searches at the LHC. We have shown that these observables (both angles and energies of the $H^\pm$ decay products) may improve the sensitivity of the LHC upon charged Higgs boson signals which can be further improved if forward-backward asymmetries are used. Further, we found out that these observables can be used for characterization analyses as post-discovery tools. We would expect that a fully-fledged signal-to-background selection should be attempted now by ATLAS and CMS, possibly in conjunction with machine learning methods trained to acquire the  spin dynamics affecting differently an $H^\pm$ induced signal and SM background. We illustrated that these kinematical variables (and the asymmetries constructed out of them) are resilient to NLO QCD corrections and the choice of the flavor scheme inside the proton. Finally, we should close by remarking that our approach can be exploited for a variety of other new physics frameworks containing such (pseudo)scalar charged states so long that they induce chiral structures in the $H^+\bar tb$ vertex that are predominantly $L$- or $R$-handed.
\section*{Acknowledgments}

The work of AJ is supported by CEPC theory program and by the National Natural Science Foundation of China 
under the Grants No. 11875189 and No.11835005.
SM is supported in part through the NExT Institute and the STFC CG ST/L000296/1.  AA  and SM acknowledge funding via the H2020-MSCA-RISE-2014 grant no. 645722 (NonMinimalHiggs). 

\bibliographystyle{JHEP}
\bibliography{biblio}

\providecommand{\href}[2]{#2}\begingroup\raggedright\begin{thebibliography}{10}

\bibitem{Abachi:1995iq}
{\bf D0} Collaboration, S.~Abachi et~al., {\it {Observation of the top quark}},
   {\em Phys. Rev. Lett.} {\bf 74} (1995) 2632--2637,
  [\href{http://arxiv.org/abs/hep-ex/9503003}{{\tt hep-ex/9503003}}].

\bibitem{Abe:1995hr}
{\bf CDF} Collaboration, F.~Abe et~al., {\it {Observation of top quark
  production in $\bar{p}p$ collisions}},  {\em Phys. Rev. Lett.} {\bf 74}
  (1995) 2626--2631, [\href{http://arxiv.org/abs/hep-ex/9503002}{{\tt
  hep-ex/9503002}}].

\bibitem{Bernreuther:2008ju}
W.~Bernreuther, {\it {Top quark physics at the LHC}},  {\em J. Phys.} {\bf G35}
  (2008) 083001, [\href{http://arxiv.org/abs/0805.1333}{{\tt
  arXiv:0805.1333}}].

\bibitem{Branco:2011iw}
G.~C. Branco, P.~M. Ferreira, L.~Lavoura, M.~N. Rebelo, M.~Sher, and J.~P.
  Silva, {\it {Theory and phenomenology of two-Higgs-doublet models}},  {\em
  Phys. Rept.} {\bf 516} (2012) 1--102,
  [\href{http://arxiv.org/abs/1106.0034}{{\tt arXiv:1106.0034}}].

\bibitem{Akeroyd:2016ymd}
A.~G. Akeroyd et~al., {\it {Prospects for charged Higgs searches at the LHC}},
  {\em Eur. Phys. J.} {\bf C77} (2017), no.~5 276,
  [\href{http://arxiv.org/abs/1607.01320}{{\tt arXiv:1607.01320}}].

\bibitem{Stange:1993td}
A.~Stange and S.~Willenbrock, {\it {Yukawa correction to top quark production
  at the Tevatron}},  {\em Phys. Rev.} {\bf D48} (1993) 2054--2061,
  [\href{http://arxiv.org/abs/hep-ph/9302291}{{\tt hep-ph/9302291}}].

\bibitem{Zhou:1996dx}
H.-Y. Zhou, C.-S. Li, and Y.-P. Kuang, {\it {Yukawa corrections to top quark
  production at the LHC in two Higgs doublet models}},  {\em Phys. Rev.} {\bf
  D55} (1997) 4412--4420, [\href{http://arxiv.org/abs/hep-ph/9603435}{{\tt
  hep-ph/9603435}}].

\bibitem{Hollik:1997hm}
W.~Hollik, W.~M. Mosle, and D.~Wackeroth, {\it {Top pair production at hadron
  colliders in nonminimal standard models}},  {\em Nucl. Phys.} {\bf B516}
  (1998) 29--54, [\href{http://arxiv.org/abs/hep-ph/9706218}{{\tt
  hep-ph/9706218}}].

\bibitem{Denner:1992vz}
A.~Denner and A.~H. Hoang, {\it {The Top decay t ---> b W in the two Higgs
  doublet model}},  {\em Nucl. Phys.} {\bf B397} (1993) 483--501.

\bibitem{Bernreuther:2008us}
W.~Bernreuther, P.~Gonzalez, and M.~Wiebusch, {\it {The Top Quark Decay Vertex
  in Standard Model Extensions}},  {\em Eur. Phys. J.} {\bf C60} (2009)
  197--211, [\href{http://arxiv.org/abs/0812.1643}{{\tt arXiv:0812.1643}}].

\bibitem{Huitu:2010ad}
K.~Huitu, S.~Kumar~Rai, K.~Rao, S.~D. Rindani, and P.~Sharma, {\it {Probing top
  charged-Higgs production using top polarization at the Large Hadron
  Collider}},  {\em JHEP} {\bf 04} (2011) 026,
  [\href{http://arxiv.org/abs/1012.0527}{{\tt arXiv:1012.0527}}].

\bibitem{Arhrib:2016vts}
A.~Arhrib and A.~Jueid, {\it {$tbW$ Anomalous Couplings in the Two Higgs
  Doublet Model}},  {\em JHEP} {\bf 08} (2016) 082,
  [\href{http://arxiv.org/abs/1606.05270}{{\tt arXiv:1606.05270}}].

\bibitem{Eilam:1990zc}
G.~Eilam, J.~L. Hewett, and A.~Soni, {\it {Rare decays of the top quark in the
  standard and two Higgs doublet models}},  {\em Phys. Rev.} {\bf D44} (1991)
  1473--1484. [Erratum: Phys. Rev.D59,039901(1999)].

\bibitem{Arhrib:2018bxc}
A.~Arhrib, A.~Jueid, and S.~Moretti, {\it {Top quark polarization as a probe of
  charged Higgs bosons}},  {\em Phys. Rev.} {\bf D98} (2018), no.~11 115006,
  [\href{http://arxiv.org/abs/1807.11306}{{\tt arXiv:1807.11306}}].

\bibitem{Aoki:2009ha}
M.~Aoki, S.~Kanemura, K.~Tsumura, and K.~Yagyu, {\it {Models of Yukawa
  interaction in the two Higgs doublet model, and their collider
  phenomenology}},  {\em Phys. Rev.} {\bf D80} (2009) 015017,
  [\href{http://arxiv.org/abs/0902.4665}{{\tt arXiv:0902.4665}}].

\bibitem{Degrande:2015vpa}
C.~Degrande, M.~Ubiali, M.~Wiesemann, and M.~Zaro, {\it {Heavy charged Higgs
  boson production at the LHC}},  {\em JHEP} {\bf 10} (2015) 145,
  [\href{http://arxiv.org/abs/1507.02549}{{\tt arXiv:1507.02549}}].

\bibitem{Guchait:2001pi}
M.~Guchait and S.~Moretti, {\it {Improving the discovery potential of charged
  Higgs bosons at Tevatron run II}},  {\em JHEP} {\bf 01} (2002) 001,
  [\href{http://arxiv.org/abs/hep-ph/0110020}{{\tt hep-ph/0110020}}].

\bibitem{Godbole:2011vw}
R.~M. Godbole, L.~Hartgring, I.~Niessen, and C.~D. White, {\it {Top
  polarisation studies in $H^-t$ and $Wt$ production}},  {\em JHEP} {\bf 01}
  (2012) 011, [\href{http://arxiv.org/abs/1111.0759}{{\tt arXiv:1111.0759}}].

\bibitem{Shelton:2008nq}
J.~Shelton, {\it {Polarized tops from new physics: signals and observables}},
  {\em Phys. Rev.} {\bf D79} (2009) 014032,
  [\href{http://arxiv.org/abs/0811.0569}{{\tt arXiv:0811.0569}}].

\bibitem{Prasath:2014mfa}
A.~Prasath~V, R.~M. Godbole, and S.~D. Rindani, {\it {Longitudinal top
  polarisation measurement and anomalous $Wtb$ coupling}},  {\em Eur. Phys. J.}
  {\bf C75} (2015), no.~9 402, [\href{http://arxiv.org/abs/1405.1264}{{\tt
  arXiv:1405.1264}}].

\bibitem{Jueid:2018wnj}
A.~Jueid, {\it {Probing anomalous $Wtb$ couplings at the LHC in single
  $t$-channel top quark production}},  {\em Phys. Rev.} {\bf D98} (2018), no.~5
  053006, [\href{http://arxiv.org/abs/1805.07763}{{\tt arXiv:1805.07763}}].

\bibitem{Alwall:2011uj}
J.~Alwall, M.~Herquet, F.~Maltoni, O.~Mattelaer, and T.~Stelzer, {\it {MadGraph
  5 : Going Beyond}},  {\em JHEP} {\bf 06} (2011) 128,
  [\href{http://arxiv.org/abs/1106.0522}{{\tt arXiv:1106.0522}}].

\bibitem{Artoisenet:2012st}
P.~Artoisenet, R.~Frederix, O.~Mattelaer, and R.~Rietkerk, {\it {Automatic
  spin-entangled decays of heavy resonances in Monte Carlo simulations}},  {\em
  JHEP} {\bf 03} (2013) 015, [\href{http://arxiv.org/abs/1212.3460}{{\tt
  arXiv:1212.3460}}].

\bibitem{Sjostrand:2014zea}
T.~Sj{\"o}strand, S.~Ask, J.~R. Christiansen, R.~Corke, N.~Desai, P.~Ilten,
  S.~Mrenna, S.~Prestel, C.~O. Rasmussen, and P.~Z. Skands, {\it {An
  Introduction to PYTHIA 8.2}},  {\em Comput. Phys. Commun.} {\bf 191} (2015)
  159--177, [\href{http://arxiv.org/abs/1410.3012}{{\tt arXiv:1410.3012}}].

\bibitem{Buckley:2010ar}
A.~Buckley, J.~Butterworth, L.~Lonnblad, D.~Grellscheid, H.~Hoeth, J.~Monk,
  H.~Schulz, and F.~Siegert, {\it {Rivet user manual}},  {\em Comput. Phys.
  Commun.} {\bf 184} (2013) 2803--2819,
  [\href{http://arxiv.org/abs/1003.0694}{{\tt arXiv:1003.0694}}].

\bibitem{Cacciari:2011ma}
M.~Cacciari, G.~P. Salam, and G.~Soyez, {\it {FastJet User Manual}},  {\em Eur.
  Phys. J.} {\bf C72} (2012) 1896, [\href{http://arxiv.org/abs/1111.6097}{{\tt
  arXiv:1111.6097}}].

\bibitem{Collaboration:2267573}
{\bf CMS Collaboration} Collaboration, C.~Collaboration, {\it {Object
  definitions for top quark analyses at the particle level}},  Tech. Rep.
  CMS-NOTE-2017-004. CERN-CMS-NOTE-2017-004, CERN, Geneva, Jun, 2017.

\bibitem{Khachatryan:2015oqa}
{\bf CMS} Collaboration, V.~Khachatryan et~al., {\it {Measurement of the
  differential cross section for top quark pair production in pp collisions at
  $\sqrt{s} = 8\,\text {TeV} $}},  {\em Eur. Phys. J.} {\bf C75} (2015), no.~11
  542, [\href{http://arxiv.org/abs/1505.04480}{{\tt arXiv:1505.04480}}].

\bibitem{Cowan:2010js}
G.~Cowan, K.~Cranmer, E.~Gross, and O.~Vitells, {\it {Asymptotic formulae for
  likelihood-based tests of new physics}},  {\em Eur. Phys. J.} {\bf C71}
  (2011) 1554, [\href{http://arxiv.org/abs/1007.1727}{{\tt arXiv:1007.1727}}].
  [Erratum: Eur. Phys. J.C73,2501(2013)].

\end{thebibliography}\endgroup

\end{document}